\documentclass[12pt]{article}

\makeatletter
\def\appendix{\par\clearpage
  \setcounter{section}{0}
  \setcounter{subsection}{0}
  \@addtoreset{equation}{section}
  \def\@sectname{Appendix~}
  \def\theequation{\thesection.\arabic{equation}}
  \def\thesection{\Alph{section}}}
\makeatother

\usepackage{amssymb}
\usepackage{graphicx}

\begin{document}

\begin{titlepage}

\phantom{.}
\vspace{-2cm}

\hfill \vbox{\hbox{DFCAL-TH 04/1}
\hbox{January 2004}}

\vskip 1.cm

\centerline{\bf Real and imaginary chemical potential in 2-color QCD}

\vskip 1.cm

\centerline{P.~Giudice and A.~Papa}

\vskip 0.5cm

\centerline{\sl Dipartimento di Fisica, Universit\`a della Calabria}
\centerline{\sl \& Istituto Nazionale di Fisica Nucleare, Gruppo collegato di 
Cosenza}
\centerline{\sl I-87036 Arcavacata di Rende, Cosenza, Italy}

\vskip 0.2cm

\centerline{{\it e-mail address:} giudice,~papa@cs.infn.it}

\vskip 1.cm

\begin{abstract}
In this paper we study the finite temperature SU(2) gauge theory with staggered 
fermions for non-zero imaginary and real chemical potential. The method of 
analytical continuation of Monte Carlo results from imaginary to real chemical 
potential is tested by comparison with simulations performed {\em directly} for 
real chemical potential. We discuss the applicability of the method in the 
different regions of the phase diagram in the temperature -- imaginary 
chemical potential plane.
\end{abstract}

\vskip 0.1cm
\hrule

\end{titlepage}
\eject
\newpage

\section{Introduction} 

Understanding the QCD phase diagram in the temperature -- chemical potential 
($T, \mu$) plane has many important implications in cosmology, in astrophysics 
and in the phenomenology of heavy-ion collisions~(for a review, see~\cite{Boy01}).
In particular, it is extremely important to localize with high accuracy the 
critical lines in this plane and to determine if the transition across them is 
first order, second order or just a crossover~\cite{Kat03}.

The formulation of the theory on a space-time lattice offers a unique tool
for such task. However, at non-zero chemical potential the 
determinant of the fermion matrix becomes complex, thus preventing to perform
standard Monte Carlo simulations. One way out is to take advantage of physical 
fluctuations in the thermal ensemble generated at $\mu=0$ in order to extract 
from it information at (small) non-zero $\mu$, after suitable 
reweighting~\cite{BMKKL98,FK02,Cro01,Eji02}. However, it is not possible 
to know {\it a priori} to what extent the reweighted ensemble overlaps the true one 
at non-zero $\mu$.\footnote{In the (2+1)-dimensional U(1) Gross-Neveu model a clear 
disagreement was evidenced~\cite{BHKLM99} in the $\mu$-dependence of the chiral 
condensate and of the number density between the determinations from 
reweighting and from standard Monte Carlo techniques. Of course, this does
not imply the ineffectiveness of the method in other theories and in different
temperature regimes.}

An alternative approach is based on the idea to perform numerical simulations 
at {\em imaginary} chemical potential~\cite{AKW99}, for which the fermion 
determinant is again real, and to analytically continue the results to real 
$\mu$~\cite{Lom00,HLP01,DLa,DLb,Pd}.\footnote{Some interesting arguments about the 
QCD phase diagram for imaginary chemical potential have been discussed in 
Ref.~\cite{Gup03}.} In practice, the method consists in fitting
the numerical data obtained for {\em imaginary} chemical potential $\mu=i\mu_I$ 
by a polynomial, which is then prolongated to {\em real} $\mu$. 
The advantage of this method lies in the fact 
that its applicability is not restricted by large lattice volumes, since it makes
no use of reweighting. Therefore, size scaling analysis can be made after the
analytical continuation to clarify the nature of the transition across the 
critical lines.
On the other side, a limitation comes from the presence of non-analyticities 
in $\mu$, thus implying, as we will see later on, that the method works 
practically only for $\mu\lesssim T$. 

In particular, if non-analyticities are singled out for imaginary $\mu$ and 
even if numerical data are obtained inside a region of analyticity in the 
$(T,\mu_I)$ plane, it is necessary that during the ``rotation'' from $\mu=i\mu_I$ 
to real $\mu=\mu_R$ no other non-analyticities are met. It is understood
that non-analyticities show up only in the infinite volume limit and that
all observables are analytic on finite lattices. However, the presence
of non-analyticities in the thermodynamical limit is indicated on (large)
finite volumes by sharp changes in the behavior of physical observables,
which is practically difficult to interpolate by polynomials, if not by
including a large number of terms. The papers where the method of analytical 
continuation has been used so far limited the attention to regions of temperature 
and chemical potential where this situation does not occur.

Of course, the compatibility between the results from two very different  
approaches, such as reweighting and analytical continuation, would give
reasonable confidence on the validity of both. Such consistency has been  
shown in the case of 4-flavor QCD in Ref.~\cite{DLa}. 

Nevertheless, a test of both methods by comparison with direct determinations
from standard Monte Carlo is interesting, especially because it 
could shed light also on the applicability outside the regions usually 
considered in literature.
For this purpose, it is necessary to turn to a theory where direct Monte  
Carlo simulations in presence of {\em real} chemical potential are feasible.

It is well known that the determinant of the fermion matrix of 2-color QCD or
SU(2) gauge theory is real even in presence of a real chemical potential (see,
for instance~\cite{AKW99}). Moreover, although being very different in the 
structure of the spectrum~\cite{HKLM99}, 2-color QCD shares many similarities
with true QCD, therefore represents the best candidate for the mentioned test.

In this paper we adopt 2-color QCD to compare determinations of the 
Polyakov loop and of the chiral condensate from the method of analytical  
continuation with results from direct Monte Carlo simulations.
In particular, we characterize different regions of the phase diagram in  
the $(T,\mu_I)$ plane according to the applicability of this method. Our findings
can be helpful to apply and to understand the results of the method of
analytical continuation in the theory of physical interest, i.e. in true QCD.

The paper is organized as follows: in Section~2 we describe the structure of the
phase diagram of 2-color QCD in the $(T,\mu_I)$ plane and discuss the 
implications on the method of analytical continuation; in Section~3 we 
present the numerical results; in Section~4 we draw some conclusions and  
anticipate some future developments. 

\section{The phase diagram of 2-color QCD}

In order to draw the phase diagram of 2-color QCD we make use of known theoretical
and numerical results on the behavior with temperature and chemical potential
of the Polyakov loop $L$ and of the chiral condensate $\langle \bar \psi \psi 
\rangle$. These objects represent true order parameters only in two limiting
cases: the Polyakov loop is the order parameter of the confinement-deconfinement
transition in the limit of infinite quark masses, while the chiral condensate is
that of the chiral transition in the limit of vanishing quark masses. However,
there is a broad numerical evidence that they both exhibit a rapid variation 
in correspondence of certain values of the parameters of the theory, which 
therefore are taken as critical ones.

\begin{figure}[tb]
\centering
\includegraphics[width=0.8\textwidth,bb=80 445 510 730,clip]{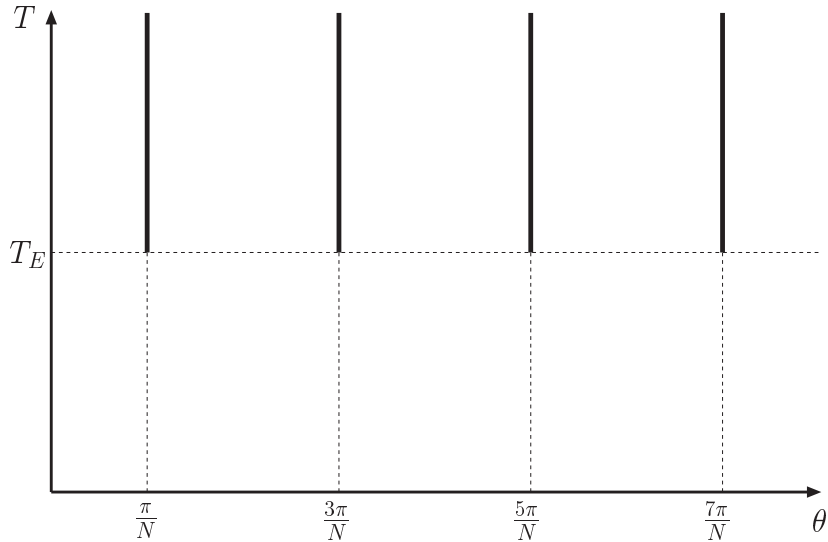}
\caption[]{Schematization of the phase diagram in the $(T,\theta)$ plane according
to Roberge and Weiss~\cite{RW86}.}
\label{fig1}
\end{figure}

We start the discussion from the case of imaginary chemical potential, $\mu=i\mu_I$.
Roberge and Weiss have shown~\cite{RW86} that the partition function of any SU(N)
theory is periodic in the parameter $\theta=\mu_I/T$ (the Boltzmann constant
is set equal to one) as consequence of a remnant of the Z(N) symmetry valid in
absence of fermions. Moreover, they have shown by weak coupling perturbation
theory that above a certain temperature $T_E$ the free energy and
the Polyakov loop show cusps or discontinuities for $\theta=2\pi(k+1/2)/N$, with
$k=0,1,\ldots$. This suggests that there are first order vertical critical lines
in the $(T,\theta)$ plane located at $\theta=2\pi(k+1/2)/N$ and extending from
$T_E$ to infinity (see Fig.~1). They argue also that below $T_E$, there is 
no phase transition for any value of $\theta$, this being true also
in the physical case of $\theta=\mu=0$.

The latter conclusion cannot be true for all values of $n_f$ and of the
fermion masses. In the chiral limit
of QCD, there are arguments~\cite{PW84} according to which a chiral phase 
transition for $\mu=0$ does exist and is first order for $n_f \geq 3$ and 
second order for $n_f=2$. For SU(2), where a strong coupling analysis could
be applied, according to Ref.~\cite{DKS86} there is a second
order chiral phase transition for $n_f=1$ and $n_f=2$. For non-vanishing masses, 
this phase transition probably becomes a crossover. Furthermore, a numerical
study on a $12^3\times 4$ lattice in SU(2) with $n_f=8$ and $am=0.07$ has 
shown a peak in the susceptibility of the chiral condensate at 
$\beta_c\simeq 1.41$, suggesting the possibility of a chiral phase transition 
near this point~\cite{LMNT00}. 

In view of all this, the worst case for the purposes of the analytical continuation
(but the most interesting from the point of view of the present study) is 
that the Roberge-Weiss (RW) 
critical line at $\theta=\pi/N$ could be prolongated downwards
and bent up to reaching the $\theta=0$ axis in correspondence of a critical 
temperature $T_c$ (see also~\cite{DLa,DLb}). For the reasons mentioned above, 
this prolongation of the RW critical line will be referred to in the following 
as the {\em chiral} critical line~\cite{DLa,DLb}. The RW periodicity and 
the symmetry for $\mu\to-\mu$, resulting
from CP invariance, imply that knowing the expectation value of an observable 
for $T$ and $\theta$ inside the strip $\{0\leq \theta \leq \pi/N, 
\; 0\leq T < \infty \}$ in the $(T,\theta)$ plane is enough to fix it
on the remaining part of the plane. Applying this argument also to the critical 
lines, we end up with the qualitative structure of the phase diagram drawn 
in Fig.~2. The actual nature of the transition across the branches of the critical 
lines of Fig.~2 which lie below $T_E$ is to a large extent unknown. 
Other scenarios of phase diagrams are possible, depending on
$n_f$ and on the fermion masses, such as, for instance, that where a chiral 
critical line emanating from $T_c$ ends somewhere before joining the vertical 
lines~\cite{Pd}.

\begin{figure}[tb]
\centering
\includegraphics[width=0.8\textwidth,bb=80 445 510 730,clip]{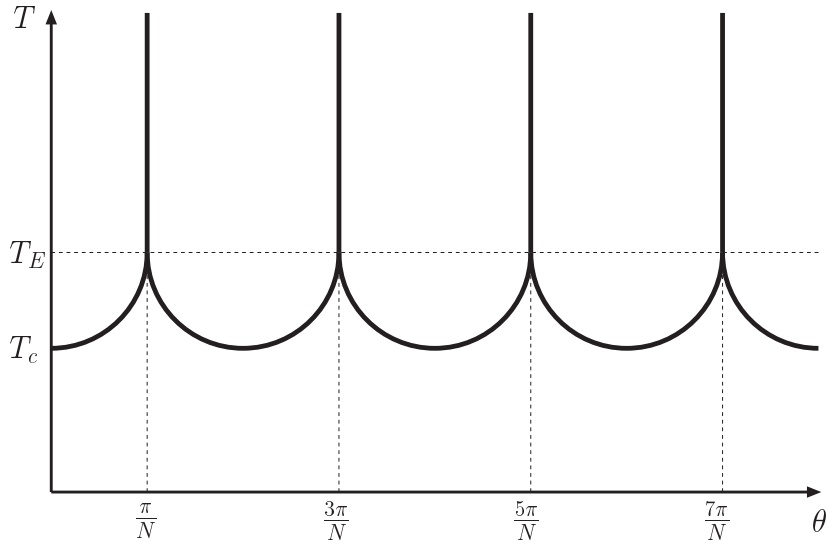}
\caption[]{Schematization of the phase diagram in the $(T,\theta)$ plane after 
the inclusion of the chiral critical lines.}
\label{fig2}
\end{figure}

Inside the mentioned strip we can single out two regions where the 
(non-perturbative) determinations of an observable can be interpolated by
a Taylor series in $\theta^2=(\mu_I/T)^2$: the region above and the region 
below the chiral critical line. This Taylor series can be analytically continued 
to real $\mu=\mu_R$ by the ``rotation'' $\mu_I \to -i \mu_R$, 
if no non-analyticities are met during this rotation. Such a condition is not 
{\it a priori} guaranteed.
It is evident, that the crucial point is how the critical line in the $(T,\theta)$ 
plane, implicitely defined by $T=f(\mu_I/T)$, evolves during the rotation 
from imaginary to real $\mu$. A reasonable assumption is that, at the end of 
the rotation, the critical line in the plane with real $\mu$ is given 
by $T=f(-i\mu_R/T)\equiv g(\mu_R/T)$, with $g$ a  
monotonically decreasing function in $\mu_R/T$. Another assumption, which needs
confirmation by numerical tests, is that the critical line evolves smoothly
during the rotation, i.e. it spans in the $(T,\mu_R/T,\mu_I/T)$ space a 
2-dimensional surface, connecting the critical line in the $(T,\mu_I/T)$ plane 
with the one in the $(T,\mu_R/T)$ plane, which does not possess bumps or 
ridges or valleys.

If these assumptions are true and the scenario of Fig.~2 holds, we can 
identify three regions in the 
strip $\{0\leq \theta \leq \pi/N, \; 0\leq T < \infty \}$ of the $(T,\mu_I/T)$ 
plane according to the possibility to apply the analytical continuation:

\begin{itemize}

\item $T>T_E$: analytical continuation possible for $0\leq \mu_I/T < \pi/N$;

\item $T_c<T<T_E$: analytical continuation possible for $0\leq \mu_I/T < \mu_I^*/T$,
where $\mu_I^*/T$ is the solution of $T=f(\mu_I^*/T)$, i.e. the value
of $\mu_I/T$ on the chiral critical line at the given temperature;

\item $T<T_c$: analytical continuation possible for $0\leq \mu_I/T < -i\mu_R^*/T$,
where $\mu_R^*/T$ is the solution of $T=g(\mu_R^*/T)$, i.e. the value
of $\mu_R/T$ on the critical line {\em of the $(T,\mu_R/T)$ plane} at 
the given temperature.

\end{itemize}

So far the method of analytical continuation has been applied, implicitely making
the assumptions we have mentioned, to the chiral condensate in the region 
$T>T_E$ for QCD with $n_f=4$~\cite{DLa} and to the critical line itself for 
QCD with $n_f=4$~\cite{DLa} and with $n_f=2,~3$~\cite{Pd}. In the case of
QCD with $n_f=4$, the analytically continued critical line has been found in 
agreement with the determination of Ref.~\cite{FK02}.

In the next Section we will compare the scenario sketched in the above
list with numerical simulations in all the three temperature regions. 
As a test-field theory we use 2-color QCD with $n_f=8$ and as observables 
we adopt the Polyakov loop $L$ and the chiral condensate $\langle \bar \psi \psi 
\rangle$.

\section{Numerical results}

We consider the formulation of 2-color QCD or SU(2) gauge theory 
on a $N_\sigma^3\times N_\tau$ lattice with staggered fermions for non-zero 
temperature and chemical potential of the baryonic density. 

The finite temperature is realized by compactifying
the time direction and by imposing (anti)periodic boundary conditions on
(fermion) boson fields. The connection between the temperature $T$ and the 
lattice size in the time direction is $T=1/(N_\tau a)$, where $a=a(\beta)$
is the lattice spacing and $\beta\equiv 4/g^2$ is the bare coupling constant. 
The chemical potential is introduced by the replacements~\cite{HK83}
\begin{eqnarray}
U_\tau &\rightarrow& e^{a\mu} \, U_\tau \nonumber \\
U_\tau^\dagger &\rightarrow& e^{-a\mu} \, U_\tau^\dagger\;,
\end{eqnarray}
where $U_\tau$ and $U_\tau^\dagger$ are the gauge link variables in the forward
and backward time direction, respectively. When $\mu$ is purely imaginary
these replacements amount to add a constant background U(1) field to the 
original theory. For SU(2) the fermionic determinant, appearing in the
partition function after integration of the fermion fields, is real for 
any complex value of the chemical potential $\mu$, thus allowing the Monte 
Carlo importance sampling without any limitation.

We adopted in this work the same hybrid simulation algorithm of Ref.~\cite{HKLM99}. 
Since for $\mu\neq 0$ it is not possible to use the even-odd partitioning, we were
forced to simulate eight degenerate continuum flavors. We performed all the 
simulations on a $8^3\times 4$ lattice, setting the quark bare mass to $am=0.07$ 
and the microcanonical time step to $dt=0.02$. We made refreshments of the 
gauge fields every 5 steps of the molecular dynamics and of the 
pseudofermion fields every 3 steps. In order to reduce autocorrelation effects, 
we made ``measurements'' every 50 steps and analyzed data by the jackknife method
combined with binning.

Before presenting the numerical results, it is convenient to redraw the phase
diagram discussed in the previous Section in units more suitable for the
lattice. In particular, on the vertical axis we put the inverse coupling
$\beta$ which is connected to the temperature $T$ by a monotonic function
in the region of interest (increasing $\beta$ means increasing temperature).
On the horizontal axis we put the chemical potential in lattice units, $\hat \mu
\equiv a \mu$. The qualitative structure of the phase diagrams 
in the $(\beta,\hat \mu_I)$ and in the $(\beta,\hat \mu_R)$ planes is the same
as in the $(T,\mu_I/T)$ and in the $(T,\mu_R/T)$ planes. In the new units, however,
the RW transition lines in the $(\beta,\hat \mu_I)$ plane appear at 
$\hat\mu_I=2\pi(k+1/2)/(N N_{\tau})$. For $N=2$ the first RW transition line
is located at $\hat \mu_I=\pi/(2 N_\tau)$. Of course, the scenario illustrated 
at the end of the previous Section must be reformulated in terms of $\beta$,
$\beta_E$ and $\beta_c$.

An estimate of the value of $\beta_c$ can be deduced from the literature. Indeed,
in Ref.~\cite{LMNT00}, a study of the chiral condensate on a $12^3 \times 4$ 
lattice for bare quark mass $am=0.07$ lead to the result $\beta_c=1.41\pm0.03$. 
Since we have a different lattice, we expect a small difference between our 
$\beta_c$ and the determination of Ref.~\cite{LMNT00}. However, for our purposes
we can safely adopt this value of $\beta_c$.

\begin{figure}[htb]
\centering
\hspace{-1.3cm}
\includegraphics[width=0.75\textwidth,angle=-90]{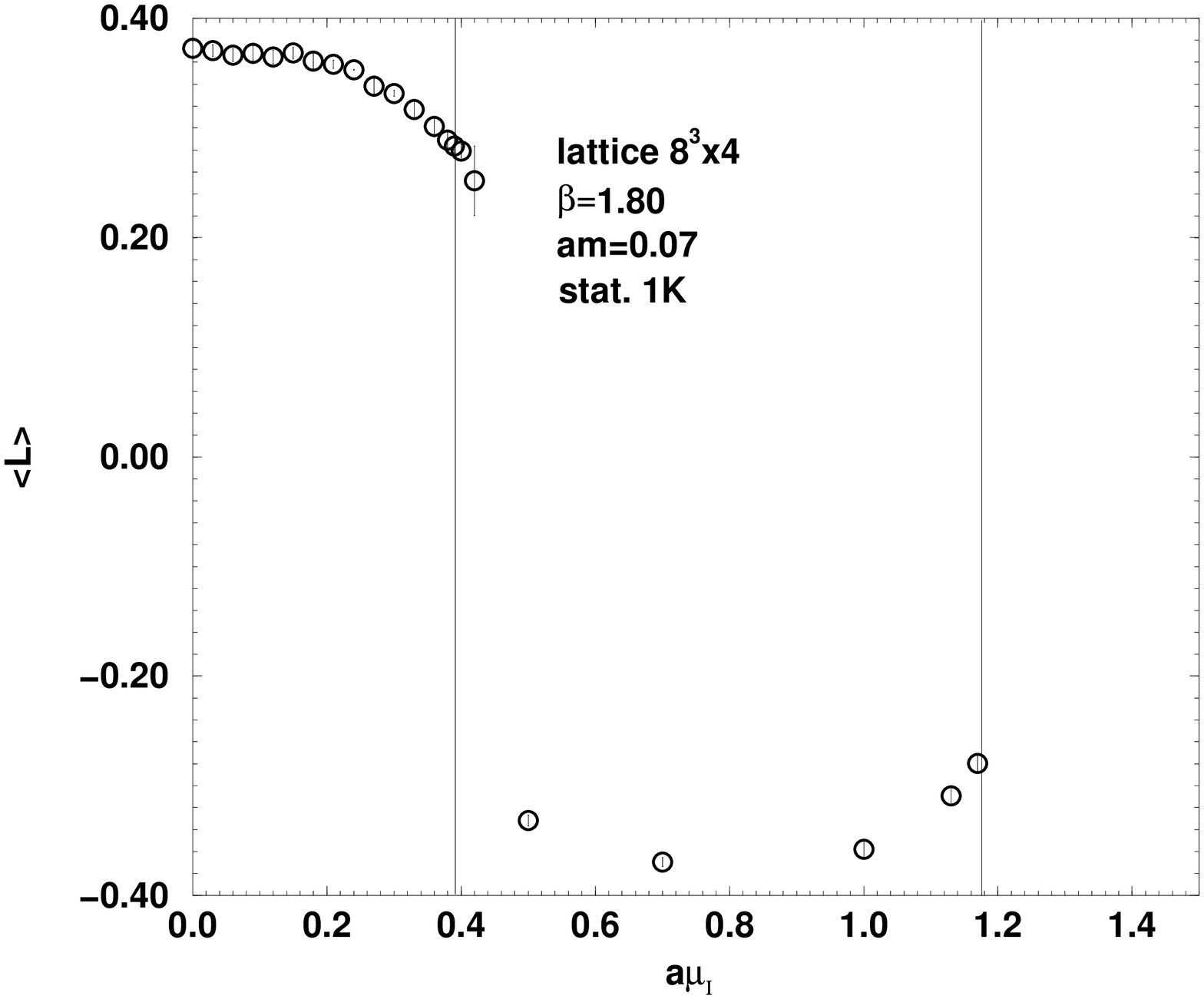}

\hspace{-1.3cm}
\includegraphics[width=0.75\textwidth,angle=-90]{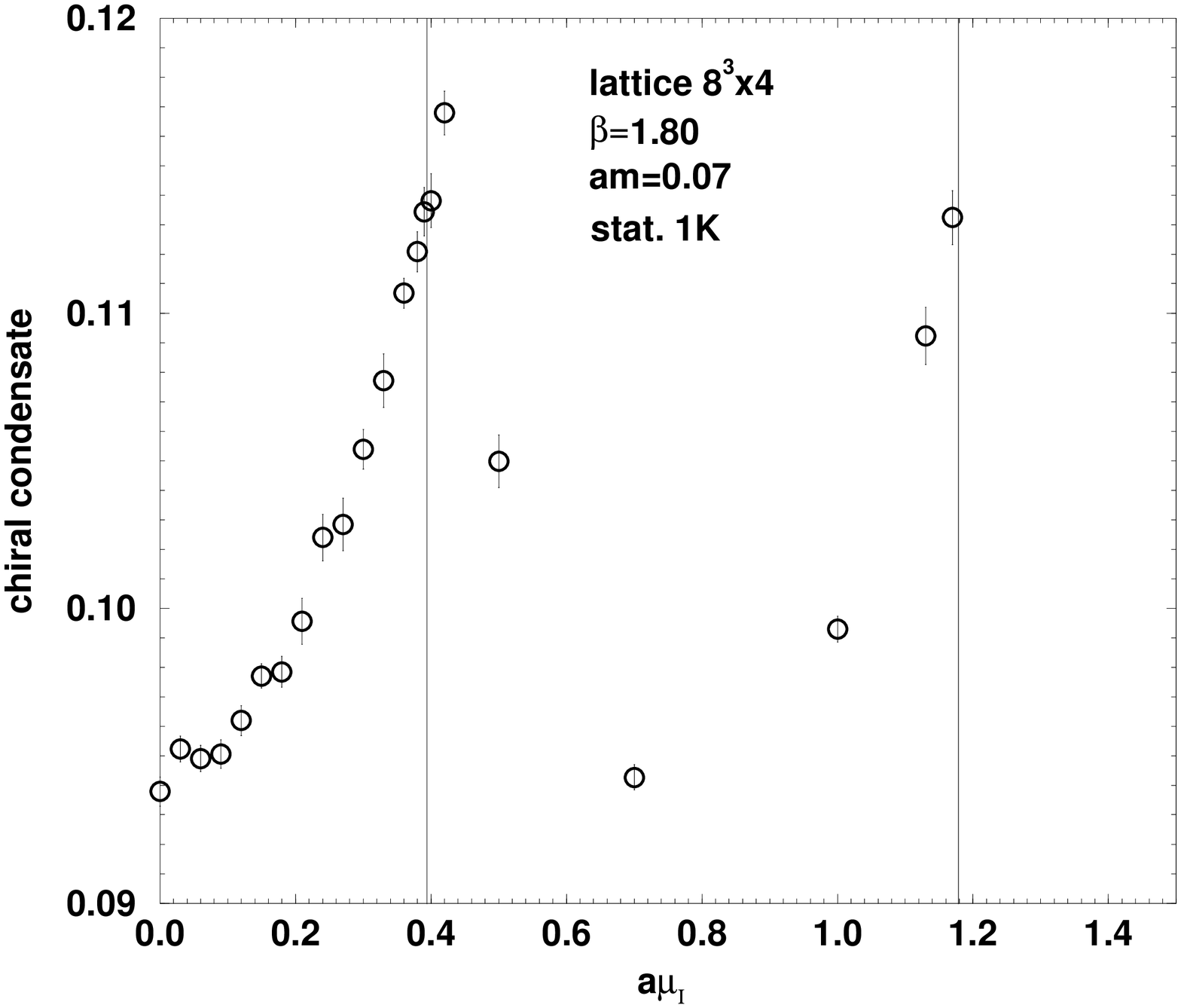}
\caption[]{Polyakov loop (above) and chiral condensate (below) {\it vs}
$\hat \mu_I$ for $\beta=1.80$ on a $8^3\times 4$ lattice. 
The solid vertical lines represent the first two RW critical lines 
($\hat\mu_I=\pi/8$ and $\hat\mu_I=3\pi/8$).}
\label{fig3}
\end{figure}

As a first step we took two $\beta$ values, one (much) above $\beta_c$ and
likely above $\beta_E$, the other below $\beta_c$ in order to possibly
single out a different behavior in $\hat\mu_I$ of lattice observables 
across $\hat\mu_I=\pi/(2N_\tau)$. We took $\beta$=1.80 for the higher $\beta$ value 
and $\beta=1.35$ for the lower and determined the Polyakov loop and the chiral 
condensate for $\hat\mu_I$ ranging between 0 and the value corresponding
to the second RW critical line, i.e. $3\pi/(2N_\tau)$. The results are shown in 
Figs.~3 and~4. We can clearly see the RW periodicity in all cases; however,
for $\beta=1.80$ we have discontinuity for the Polyakov loop and cusp for the 
chiral condensate across the first RW critical line, while for $\beta=1.35$ 
the behavior is smooth for both observables. We have verified that the overshooting
of the first RW critical line seen in Fig.~3 is due to an hysteresis effect, which
confirms the first order nature of the transition. 

\begin{figure}[htb]
\centering
\hspace{-1.3cm}
\includegraphics[width=0.75\textwidth,angle=-90]{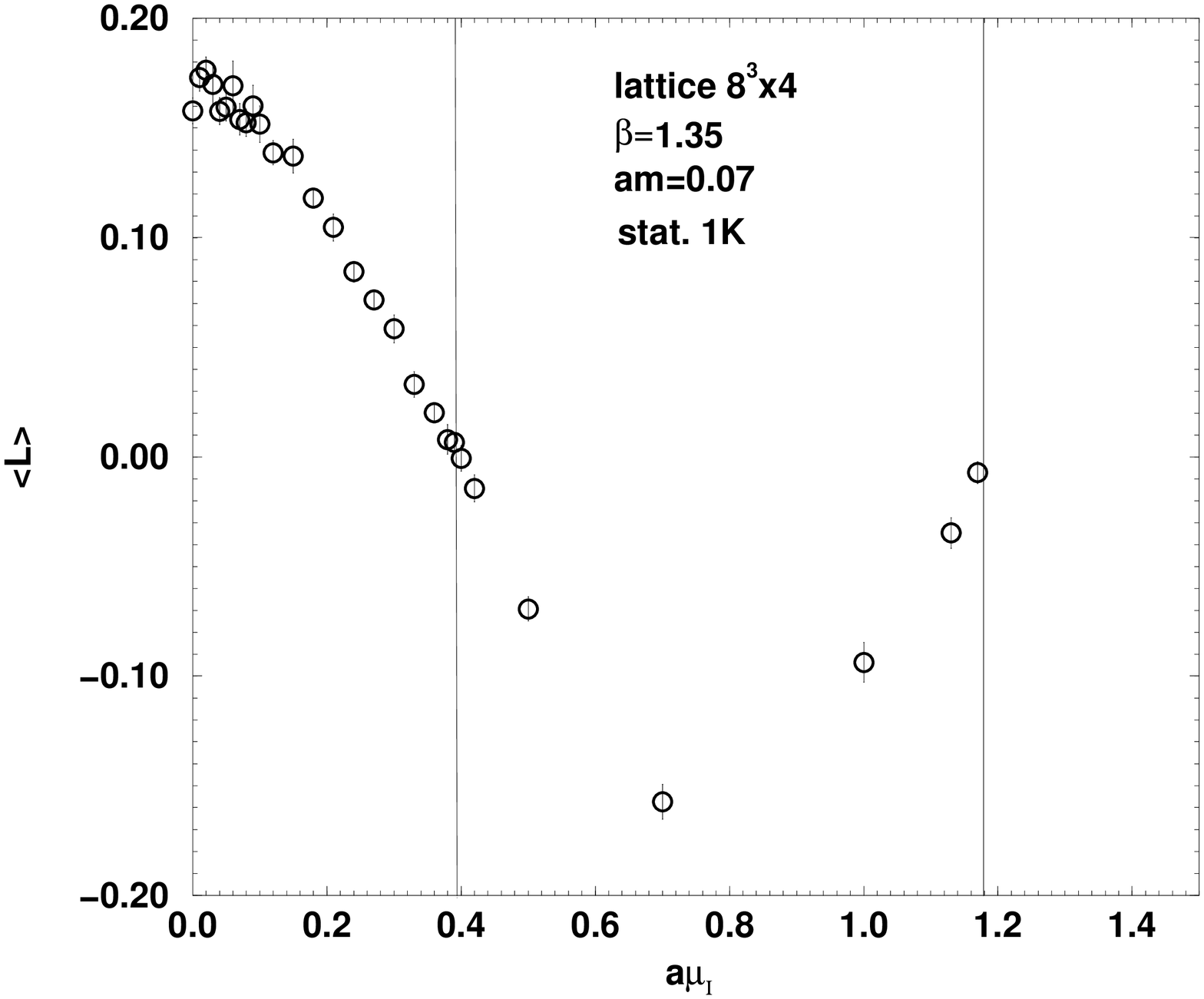}

\hspace{-1.3cm}
\includegraphics[width=0.75\textwidth,angle=-90]{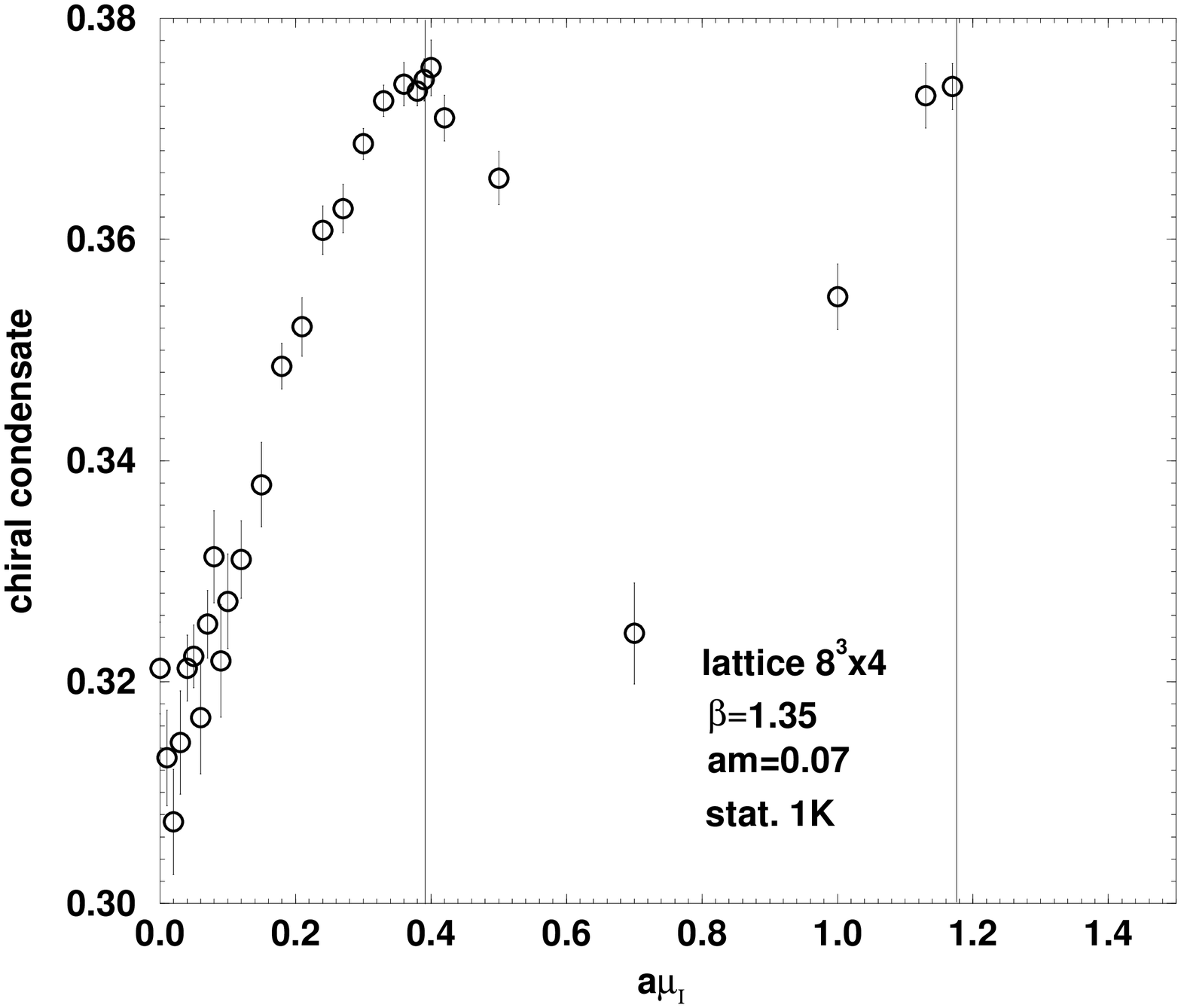}
\caption[]{As in Fig.~3 for $\beta=1.35$.}
\label{fig4}
\end{figure}

In order to proceed with the test of the method of analytical continuation
we need also an estimation of $\beta_E$. There is a nice procedure to do it,
suggested in Ref.~\cite{DLa}. It relies on performing simulations at a critical
value of $\hat \mu_I$, say $\hat\mu_I=\pi/(2N_\tau)$, and on changing $\beta$ from a
very high value downwards. If the zero field configuration is chosen as the 
starting one, the system remains always in the phase on the left of the RW line
until $\beta>\beta_E$, if the flip to the other phase is prevented by a large 
space volume. Due to our limited computational power, we are constrained to 
a space volume which is too small to prevent flips between the two phases.
An alternative procedure to determine $\beta_E$ could be to study the behavior 
of the Polyakov loop (or of its phase) and of the chiral condensate across the 
RW critical lines for different $\beta$ values and to see when the 
discontinuities or the cusps are smoothed out. Another possibility could be to
exploit the hysteresis effect in the Polyakov loop or in the 
chiral condensate determinations across the RW critical line. 
The procedure consists in building two sequences of expectation values 
of the Polyakov loop; in the first (second) sequence the starting configuration 
of any new run is taken to be a thermal equilibrium configuration being 
always in the phase on the left (right) side of the first RW transition. Across 
the RW critical line the comparison of these two sequences of data shows 
a hysteresis effect as long as $\beta>\beta_E$.
This effect disappears below $\beta_E$. Of course, both the above procedures 
demand for a high statistics and, so far, we can only say that $\beta_E$ is bounded 
between about 1.53 and about 1.57. This level of accuracy is, however, enough 
for the purposes of the present study.

We turn now to the main part of this work, i.e. the test if the scenario
presented at the end of the previous Section holds. We selected three values of
$\beta$ (1.90, 1.45, 0.90), each representative of one of the regions 
$\beta>\beta_E$, $\beta_c<\beta<\beta_E$ and $\beta<\beta_c$. For each $\beta$
we have performed simulations for both real and imaginary chemical
potential, varying $\hat \mu_I$ and $\hat \mu_R$ between 0 and $\pi/(2 N_\tau)$
and have determined expectation values of the Polyakov loop and of the 
chiral condensate. Then, we have interpolated the data obtained for imaginary $\mu$
with a {\em truncated} Taylor series of the form $a+b\,\hat \mu_I^2+c\,\hat
\mu_I^4$. 
After ``rotating'' this polynomial to real chemical potential, thus leading to
$a-b\,\hat\mu_R^2+c\,\hat\mu_R^4$, we have compared it with the 
determinations obtained directly for real $\mu=\mu_R$.

For $\beta=1.90$ we have interpolated with the fourth order polynomial
the data obtained for $0\leq\hat\mu_I\lesssim \pi/(2N_\tau)$
and have found that the rotated polynomial perfectly interpolates
data obtained directly for real $\mu=\mu_R$ in the same range (see Fig.~5). 
The uncertainty in the rotated polynomial has been obtained
by propagating in standard way the uncertainty on the fitted coefficients $a$,
$b$ and $c$. We have verified that the same situation occurs for other values of
$\beta$ larger than $\beta_E$. From this outcome we can conclude that 
the analytical continuation works in the region $\beta>\beta_E$ for all the
$\hat\mu_I$'s before the first RW critical line.

\begin{figure}[htb]
\centering
\hspace{-1.3cm}
\includegraphics[width=0.75\textwidth,angle=-90]{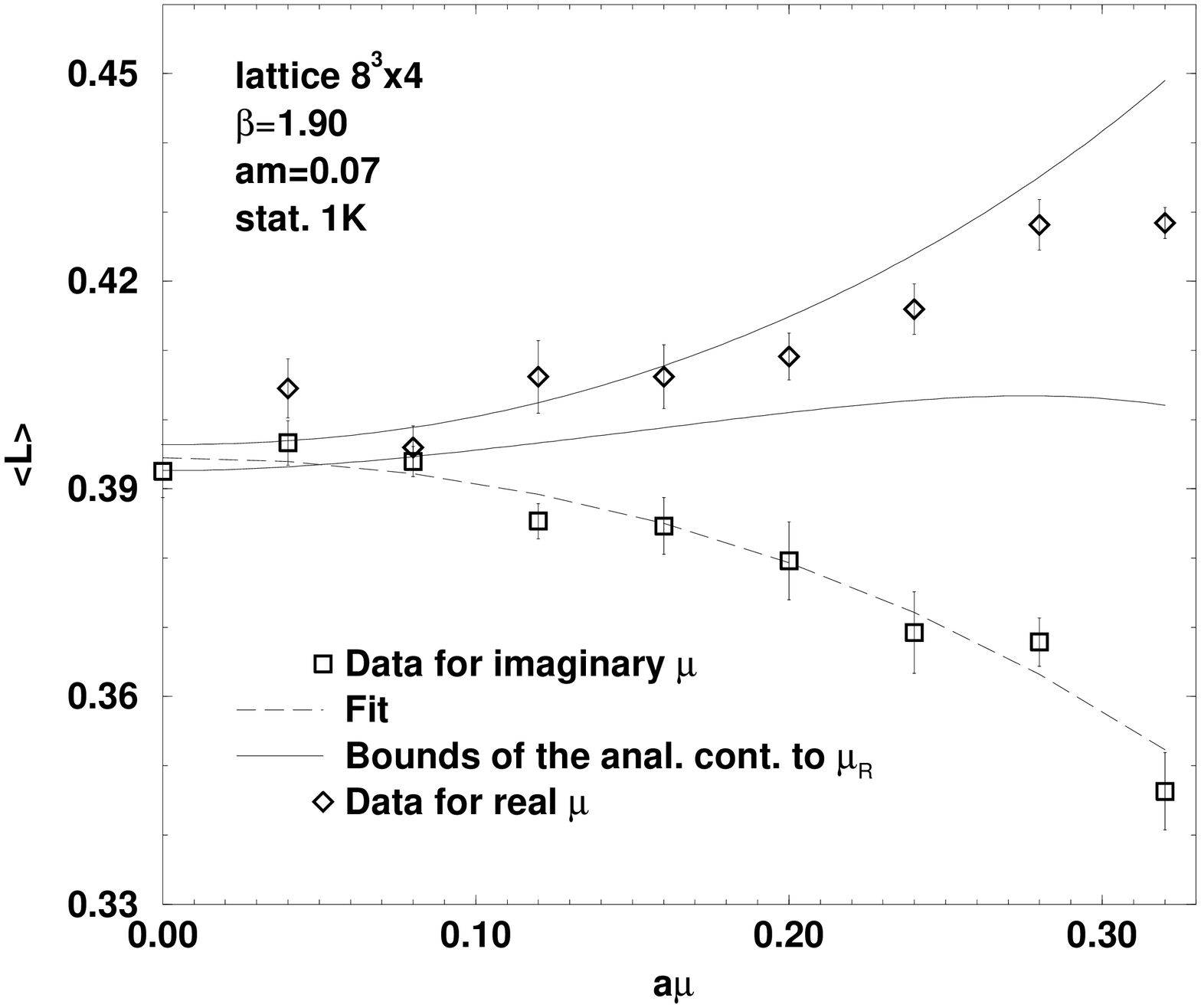}

\hspace{-1.3cm}
\includegraphics[width=0.75\textwidth,angle=-90]{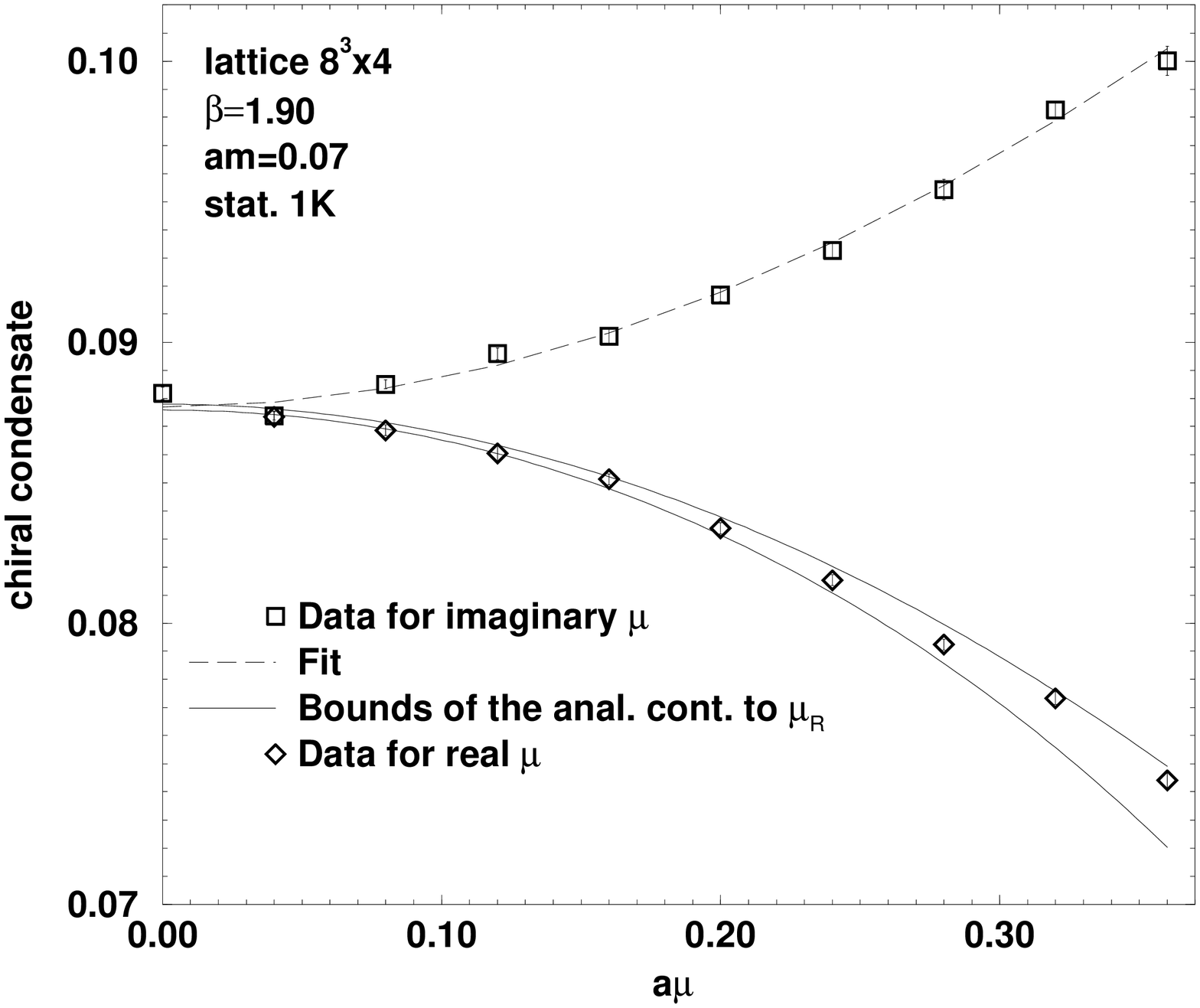}
\caption[]{Polyakov loop (above) and chiral condensate (below) {\it vs}
$\hat \mu$ for $\beta=1.90$ on a $8^3\times 4$ lattice. Numerical data
for imaginary chemical potential (squares) are fitted by a fourth order polynomial
(long-dashed line). The ``rotated'' polynomial, whose upper and lower bounds
are given by the solid lines, is compared with data obtained for real chemical 
potential (diamonds).}
\label{fig5}
\end{figure}

For $\beta=1.45$ we are in the region $\beta_c < \beta <\beta_E$ for which,
if the scenario presented at the end of the previous Section holds, we 
expect that by varying $\hat \mu_I$ we meet the chiral critical line, while
no transitions should be met at the same $\beta$ by varying $\hat \mu_R$.
Therefore this time we have interpolated with the fourth order polynomial 
the data obtained for the {\em real} chemical potential $0\leq\hat\mu_R
\lesssim\pi/(2N_\tau)$ and have rotated this polynomial to its
counterpart in $\hat\mu_I$. The comparison with data obtained directly 
for imaginary chemical potential shows agreement for $\hat\mu_I$ below
$\simeq 0.18$, which can be taken as an estimate of the chiral critical value 
in the $(\beta,\hat\mu_I)$ plane at the given $\beta$ (see Fig.~6). 

\begin{figure}[htb]
\centering
\hspace{-1.3cm}
\includegraphics[width=0.75\textwidth,angle=-90]{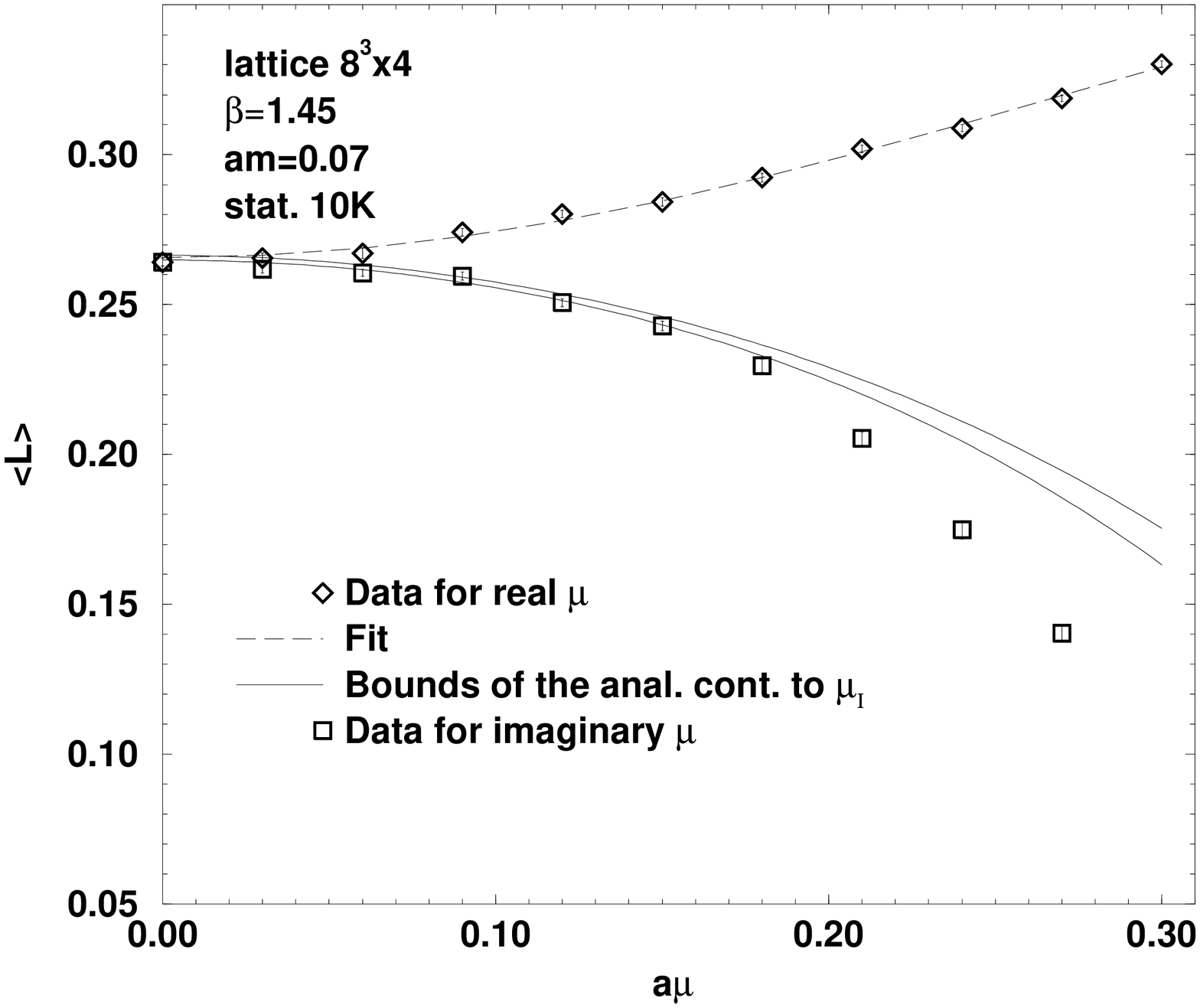}

\hspace{-1.3cm}
\includegraphics[width=0.75\textwidth,angle=-90]{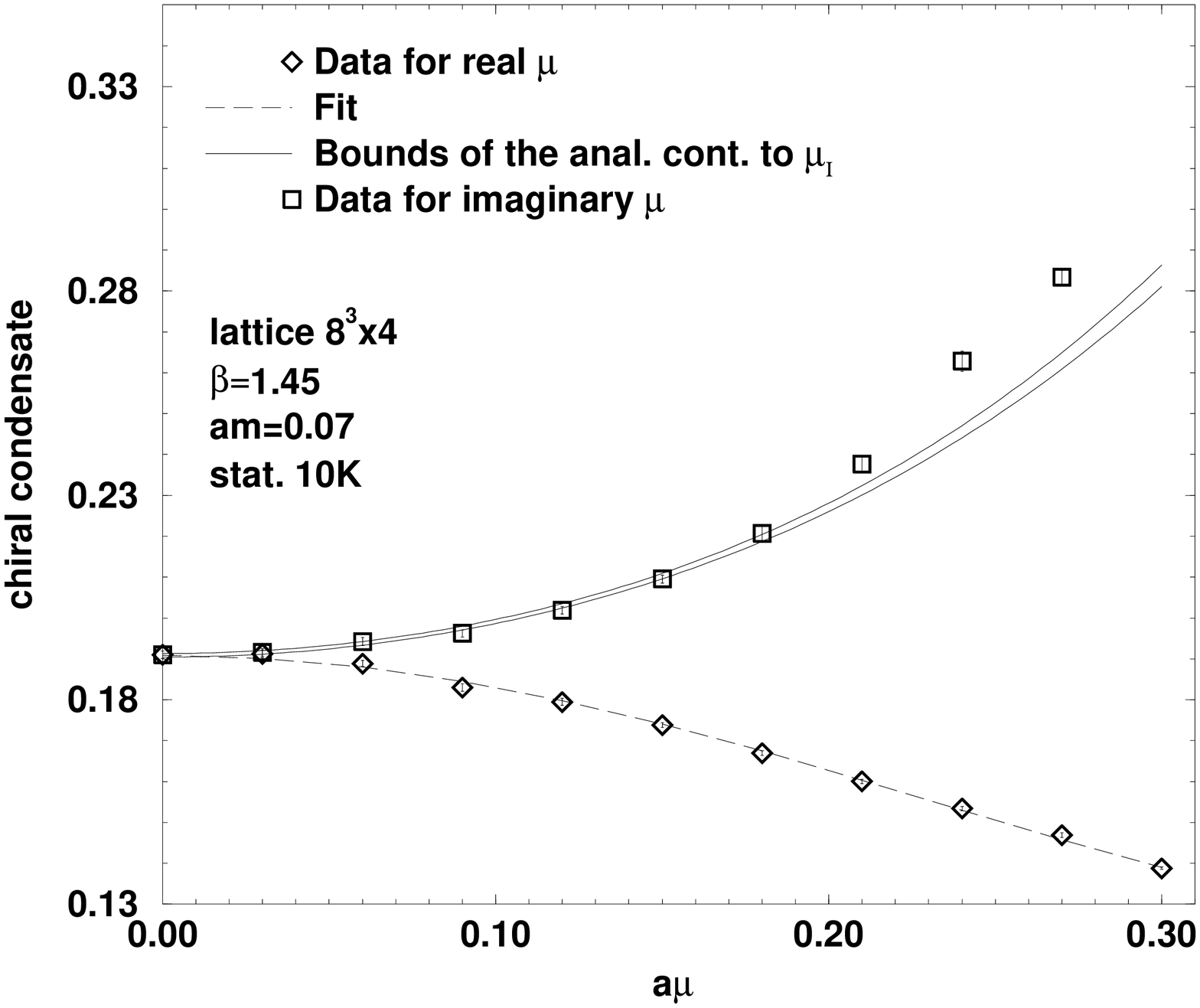}
\caption[]{Polyakov loop (above) and chiral condensate (below) {\it vs}
$\hat \mu$ for $\beta=1.45$ on a $8^3\times 4$ lattice. Numerical data
for real chemical potential (diamonds) are fitted by a fourth order polynomial
(long-dashed line). The ``rotated'' polynomial, whose upper and lower bounds
are given by the solid lines, is compared with data obtained for imaginary chemical 
potential (squares).}
\label{fig6}
\end{figure}

For $\beta=0.90<\beta_c$, we expect analyticity for all possible
values of $\hat \mu_I$, while the chiral critical line in the $(\beta,\hat\mu_R)$
plane is met by varying $\hat\mu_R$ at the given value of $\beta$.   
Therefore we have interpolated with the fourth order polynomial 
the data obtained for $0\leq\hat\mu_I\lesssim\pi/(2N_\tau)$ and 
have rotated this polynomial to its counterpart in $\hat\mu_R$. The comparison 
with data obtained directly for real chemical potential shows agreement for 
$\hat\mu_R$ below $\simeq 0.12$, which can be taken as an estimate of the critical
value in the $(\beta,\hat\mu_R)$ plane at the given $\beta$ (see Fig.~7). 

\begin{figure}[htb]
\centering
\hspace{-1.3cm}
\includegraphics[width=0.75\textwidth,angle=-90]{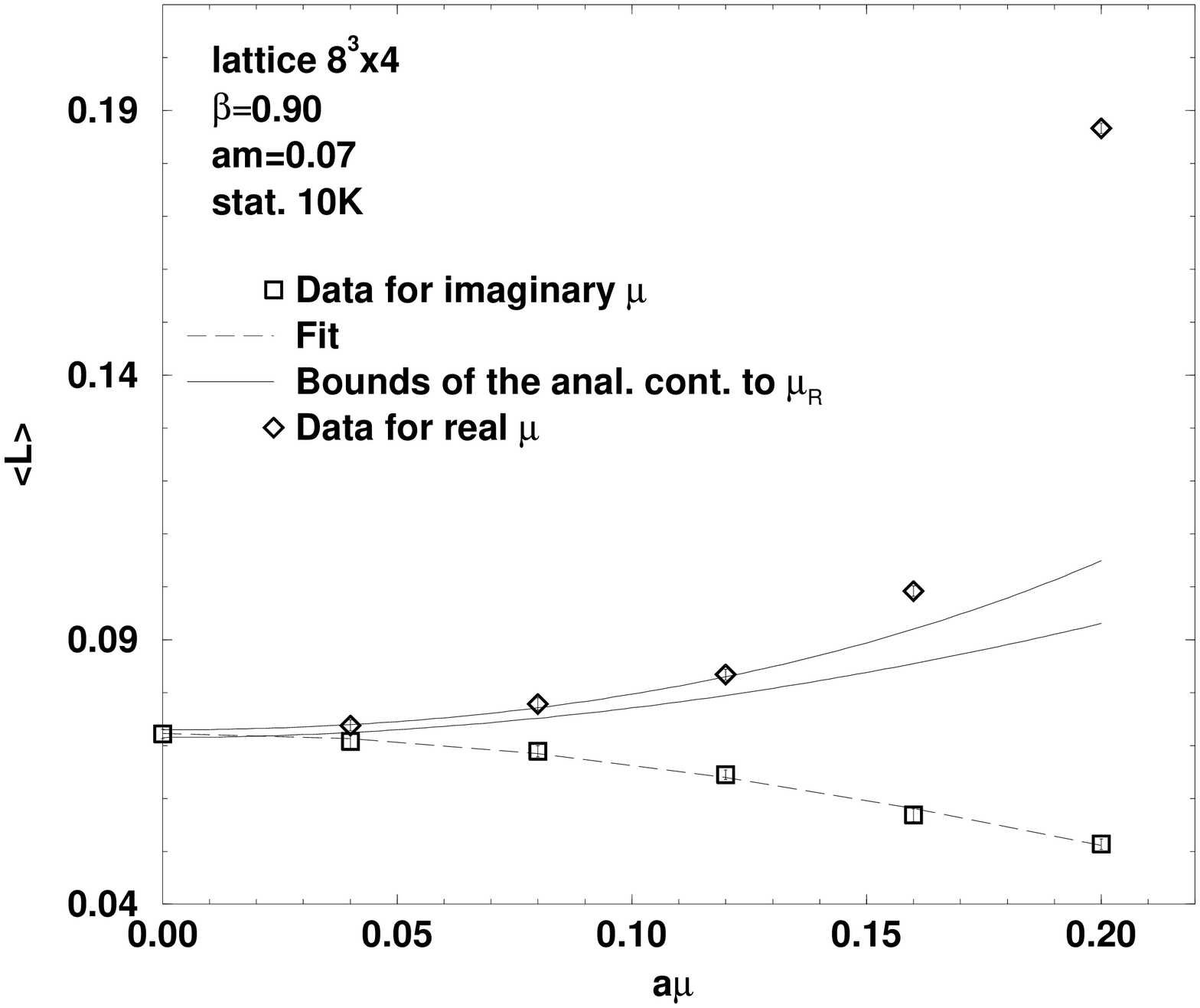}

\hspace{-1.3cm}
\includegraphics[width=0.75\textwidth,angle=-90]{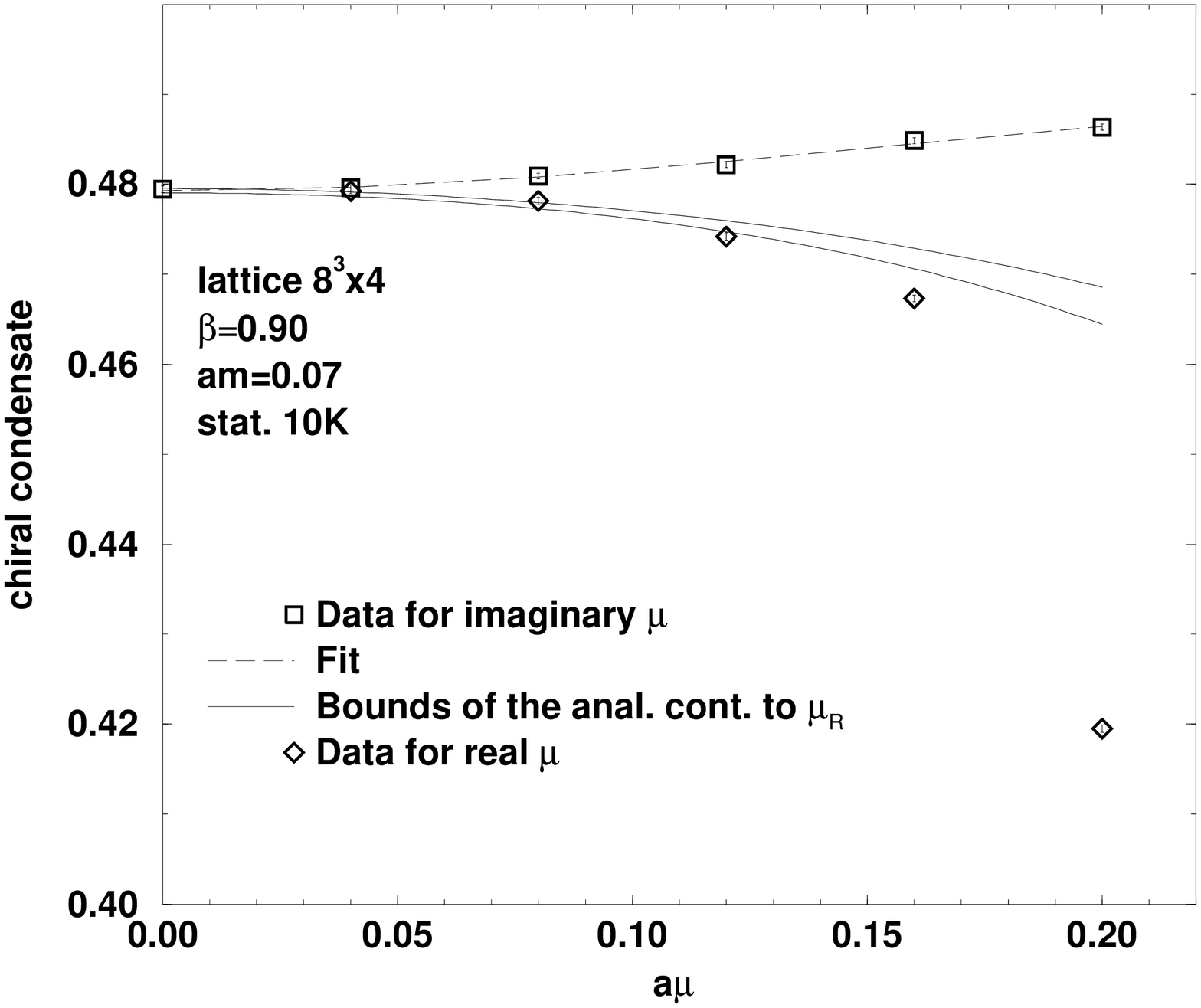}
\caption[]{As in Fig.~5 for $\beta=0.90$.}
\label{fig7}
\end{figure}

We can conclude from these results that the scenario of the end of the previous 
Section is consistent, thus giving support to the assumptions underlying it.

\section{Conclusions and outlook}

In this paper we have presented a possible structure of the phase diagram
in the temperature -- imaginary chemical potential in 2-color QCD and have
argued how this structure can be continued to real chemical potential.
Then, we have shown that the scenario we have described is consistent with
Monte Carlo numerical determinations of the Polyakov loop and of the chiral
condensate for real and imaginary chemical potential.

In particular, we have found that the method of analytical continuation, largely
exploited now in the literature for the physically interesting case of QCD, 
works fine within the restrictions posed by the presence of non-analyticities.

High precision determinations were not the aim of this paper. We limited 
ourselves to give convincing numerical support to the proposed scenario,
postponing massive calculations to forthcoming works.


We plan, moreover, to further develop the present ideas by (a) determining
the location of the critical lines for real and imaginary chemical potential
by the standard method of susceptibilities, (b) verifying that the critical
lines for real chemical potential can be obtained from analytical
continuation of those for imaginary chemical potential, (c) following the 
``evolution'' of the critical lines during the rotation from imaginary to real 
chemical potential by simulations with {\em complex} chemical potential.

\vspace{0.5cm} {\bf \large Acknowledgments} We have the pleasure to thank 
M.-P.~Lombardo for many stimulating discussions and for providing us with the Monte 
Carlo code. We acknowledge also several interesting conversations with 
R.~Fiore.

\vfill \eject

\end{document}